\documentstyle[prd,preprint,tighten,aps,floats,eqsecnum,%
amssymb,amsfonts,newlfont,graphicx]{revtex}
\begin{document}
\draft
\preprint{
\begin{tabular}{r}
   DFPD 00/EP/51
\\ DFTT 40/00
\\ arXiv:hep-ex/0011069
\end{tabular}
}
\title{The Power of Confidence Intervals}
\author{Carlo Giunti}
\address{INFN, Sez. di Torino, and Dip. di Fisica Teorica,
Univ. di Torino, I--10125 Torino, Italy}
\author{Marco Laveder}
\address{Dip. di Fisica ``G. Galilei'', Univ. di Padova,
and INFN, Sez. di Padova, I--35131 Padova, Italy}
\date{21 November 2000}
\maketitle
\begin{abstract}
We consider the power to reject false values of the parameter
in Frequentist methods
for the calculation of confidence intervals.
We connect the power with the physical significance
(reliability)
of confidence intervals
for a parameter bounded to be non-negative.
We show that
the confidence intervals
(upper limits) obtained with a (biased) method that
near the boundary has
large power in testing the parameter against larger alternatives
and
small power in testing the parameter against smaller alternatives
are physically more significant.
Considering the recently proposed methods with correct coverage,
we show that
the physical significance of
upper limits is smallest in the Unified Approach
and highest in the Maximum Likelihood Estimator method.
We illustrate our arguments in
the specific cases of a bounded Gaussian distribution
and a Poisson distribution with known background.
\end{abstract}

\section{Introduction}
\label{Introduction}

The problem of calculating Frequentist confidence intervals
and upper limits
has received recently new contributions
\cite{Feldman-Cousins-98,Ciampolillo-98,%
Giunti-bo-99,Giunti-back-99,%
Roe-Woodroofe-99,%
Mandelkern-Schultz-99,Punzi-99,Narsky-99,Cousins-00,%
Giunti-Laveder-significance-00,Roe-Woodroofe-00}
and has been subject of intense discussions
(see Refs.~\cite{CERN-CLW-2000,FNAL-CLW-2000,Giunti-use-00}).
Three new methods with correct coverage
(see Section~\ref{Coverage and power})
have been proposed:
Unified Approach \cite{Feldman-Cousins-98},
Bayesian Ordering \cite{Giunti-bo-99}
and
Maximum Likelihood Estimator\footnote{
We introduce here this name for the method proposed by
Mandelkern and Schultz in Ref.~\cite{Mandelkern-Schultz-99},
since it is based on the use of the estimator
derived from maximum likelihood.
}\cite{Mandelkern-Schultz-99}.

In this paper we consider
the power of Frequentist methods
to reject false values of the parameter under investigation
and we connect it with the physical significance of confidence intervals.
In this contest the
physical significance of a confidence interval
is its degree of reliability.
For example, an unbelievably small
upper limit
below the sensitivity of the experiment
has negligible physical significance
(one could even argue that it has ``negative''
physical significance,
since it gives misleading information
to those who believe in it).
Empty confidence intervals are practically
useless and physically insignificant.

In Section~\ref{Coverage and power}
we review the coverage of Frequentist methods
and their power
to reject false values of the parameter.
Section~\ref{boundary}
constitutes the main part of the paper,
in which considering a non-negative parameter,
we discuss the power of confidence intervals
and their physical significance.
In Sections~\ref{Gaussian} and \ref{Poisson}
we illustrate the arguments
presented in Section~\ref{boundary}
using as examples a Gaussian distribution with mean $\mu\geq0$
and a Poisson distribution with known background,
respectively.

\section{Coverage and power}
\label{Coverage and power}

An important property of
Frequentist confidence intervals
is \emph{coverage}.
A method for the calculation of confidence intervals has correct coverage
if its confidence intervals with $100(1-\alpha)\%$ confidence level (CL)
belong
to a set of confidence intervals
that can be obtained with a large ensemble of experiments,
$100(1-\alpha)\%$ of which contain the true value of the parameter
(see, for example,
Refs.~\cite{Eadie-71,Kendall-2A,Cousins-95,Giunti-back-99}).
In other words,
if coverage is satisfied,
a $100(1-\alpha)\%$ CL confidence interval
has a probability
$1-\alpha$ to cover the true value of the parameter.

Confidence intervals with correct coverage can be calculated using
Neyman's method
\cite{Neyman-37}.
In this method,
$100(1-\alpha)\%$ CL confidence intervals
for a quantity $\mu$
are obtained through the construction of
a confidence belt in the plane
$\hat\mu$--$\mu$,
where $\hat\mu$ is an appropriate estimator of $\mu$.
For each possible value of $\mu$
one calculates an acceptance interval of the estimator
with integral probability $1-\alpha$.
The union of all the acceptance intervals
constitutes the confidence belt.
The confidence interval resulting from a measurement of
$\hat\mu$
is given by all the values of $\mu$
whose acceptance interval include the measured value of $\hat\mu$
(see, for example, Refs.~\cite{Eadie-71,Kendall-2A}).

Coverage, however, is not the only quantity that is
important in the construction of confidence intervals.
Another quantity, called \emph{power},
is related to the probability
to reject false values of the parameter.
Coverage and power are connected, respectively, with the so-called
\emph{Type I} and \emph{Type II} errors in testing a
simple statistical hypothesis $H_0$
against a simple alternative hypothesis $H_1$
(see Ref.~\cite{Kendall-2A}, section 20.9):
\begin{description}
\item[Type I error:]
Reject the null hypothesis $H_0$ when it is true.
The probability of a Type I error is called
\emph{size} of the test and it is usually denoted by $\alpha$.
\item[Type II error:]
Accept the null hypothesis $H_0$ when the alternative hypothesis $H_1$ is true.
The probability of a Type II error is usually denoted by $\beta$.
The power of a test is the probability $\pi=1-\beta$
to reject $H_0$ if $H_1$ is true.
A test is \emph{most powerful}
if its power
is the largest one among all possible tests.
\end{description}

In Neyman's method,
whatever is the true value of $\mu$,
the probability that it is not included in a
$100(1-\alpha)\%$ CL confidence interval
is $\alpha$.
From the point of view of hypothesis testing,
if one considers a possible value $\mu_0$ of $\mu$
as a null hypothesis,
$H_0$: $\mu=\mu_0$,
the probability to reject $\mu_0$ if it is true is $\alpha$,
\textit{i.e.}
there is a probability $\alpha$ to make a Type I error.
In other words,
the acceptance interval of the estimator $\hat\mu$
corresponding to $\mu_0$
is the \emph{acceptance region} of the test
and the complementary interval
is the \emph{critical region} of the test.
If the measured value of $\hat\mu$
falls in the critical region,
the null hypothesis is rejected.
This happens with a probability $\alpha$ if the null hypothesis is true,
as required by coverage.

The property of coverage
is not sufficient to specify uniquely how to construct the confidence
belt.
Different Frequentist methods with correct coverage
follow different prescriptions
for the definition of the acceptance intervals.
The associated probability $\alpha$ of a Type I error is the same,
but the probability $\beta$ of a Type II error and the corresponding power
$\pi=1-\beta$
are different.

Unfortunately,
the power associated with a confidence belt
is not easy to evaluate,
because for each possible value $\mu_0$ of $\mu$
considered as a null hypothesis
there is no simple alternative hypothesis
that allows to calculate the probability $\beta$
of a Type II error.
Instead,
we have the alternative hypothesis
$H_1$: $\mu_1\neq\mu_0$,
which is composite.
For each value of $\mu_1\neq\mu_0$
one can calculate the probability $\beta_{\mu_0}(\mu_1)$
of a Type II error associated with a given acceptance interval
corresponding to $\mu_0$.
A method that gives an acceptance region
for $\mu_0$
which has the largest possible power
$\pi_{\mu_0}(\mu_1)=1-\beta_{\mu_0}(\mu_1)$
if $\mu_1$ is true
is most powerful with respect to the alternative $\mu_1$.
Clearly,
it would be desirable to find a
\emph{uniformly most powerful}
test,
\textit{i.e.}
a test that gives an acceptance region
for $\mu_0$
which has the largest possible power
$\pi_{\mu_0}(\mu_1)$
for any value of $\mu_1$.
Unfortunately,
the Neyman-Pearson lemma implies that
in general a uniformly most powerful test
does not exist if the alternative hypothesis is
\emph{two-sided},
\textit{i.e.}
both $\mu_1<\mu_0$ and $\mu_1>\mu_0$ are possible,
and the derivative of the Likelihood with respect to $\mu$
is continuous in $\mu_0$
(see Ref.~\cite{Kendall-2A}, section 20.18).
Nevertheless,
it is possible to find a uniformly most powerful test
if the class of tests is restricted in appropriate ways.
A class of tests that has some merit
is the class of \emph{unbiased} tests,
such that
\begin{equation}
\inf_{\mu_1} \pi_{\mu_0}(\mu_1)
\geq
\alpha
\,,
\label{unbiased}
\end{equation}
\textit{i.e.}
the probability of rejecting $\mu_0$ when it is false
is at least as large as the probability of
rejecting $\mu_0$ when it is true.
The \emph{equal-tail} test
used in the Central Intervals method
is unbiased
and \emph{uniformly most powerful unbiased}
for distributions belonging to the exponential family,
such as, for example,
the Gaussian and Poisson distributions
(see Ref.~\cite{Kendall-2A}, section 21.31).

Therefore,
the Central Intervals method,
is widely used because it corresponds to
a uniformly most powerful unbiased test.
Other methods based on asymmetric tests unavoidably introduce
some bias.

\section{Power near a boundary}
\label{boundary}

In some cases
the Central Intervals method is not satisfactory.
The two cases which occur often in physics are
the Gaussian distribution with a mean $\mu$ physically bounded
to be non-negative\footnote{
In general one can consider any boundary for $\mu$.
We consider only the case $\mu\geq0$ for simplicity.
}
and the Poisson distribution with mean $\mu\geq0$ and a known background.
In these cases the Central Intervals method
sometimes produces empty confidence intervals,
that are physically useless.
The recently proposed Frequentist methods with correct coverage
\cite{Feldman-Cousins-98,Giunti-bo-99,Mandelkern-Schultz-99}
cure this problem considering appropriate
constructions of the confidence belt
that guarantee a
transition from two-sided confidence intervals to upper limits
near the boundary for the Gaussian distribution
and for small number of counts
in the case of the Poisson distribution.
From the discussion at the end of Section~\ref{Coverage and power},
it is clear that these methods are biased.
In particular,
since the acceptance intervals shift towards lower values
of the estimator, which are more likely
for smallest values of $\mu$,
these methods have a bias in testing the alternatives $\mu_1<\mu_0$.
On the other hand,
these methods are more powerful than
the Central Intervals method
in the test of the alternatives $\mu_1>\mu_0$.

Near the boundary
$\mu\geq0$,
where the various methods produce different results,
it is clearly much more important to
test the alternatives $\mu_1>\mu_0$
than the alternatives $\mu_1<\mu_0$,
which are limited.
In other words,
the boundary introduces an asymmetry
in the importance of the
$\mu_1>\mu_0$
and
$\mu_1<\mu_0$
alternatives.
Moreover,
experiments are made to search for a signal
and
when small signals are searched for
(small $\mu$ near the boundary $\mu\geq0$),
testing the alternatives $\mu_1>\mu_0$
is physically more meaningful
than
testing the alternatives $\mu_1<\mu_0$.
Having more power
in the test of the alternatives $\mu_1>\mu_0$
means that values of $\mu$ smaller than the true one
are less likely to be accepted,
\textit{i.e.}
the experiment is more sensitive to a possible signal.

In general the loss of power in testing
$\mu_0$ against $\mu_1<\mu_0$
when $\mu_0$ is near the physical boundary
leads to less stringent upper limits.
If the experiment is not sensitive to values of $\mu$
near the physical boundary,
it does not make any sense
to test $\mu_0$ against $\mu_1<\mu_0$
near the boundary.
Hence,
a loss of power for this test is actually desirable
and leads to more reliable upper limits.
Indeed,
the Central Intervals method,
which is unbiased and has a large power in testing
$\mu_0$ against $\mu_1<\mu_0$
near the boundary,
gives practically useless upper limits.

These considerations imply that a Frequentist method produces
upper limits which are
\emph{physically significant}
if near the physical boundary it has a
\emph{large} power in testing
$\mu_0$ against $\mu_1>\mu_0$
and \emph{small} power in testing
$\mu_0$ against $\mu_1<\mu_0$.

Among the recently proposed Frequentist methods
with correct coverage
(Unified Approach \cite{Feldman-Cousins-98},
Bayesian Ordering \cite{Giunti-bo-99},
Maximum Likelihood Estimator \cite{Mandelkern-Schultz-99}),
the shift of the acceptance intervals towards lower values
of the estimator is smallest in
the Unified Approach \cite{Feldman-Cousins-98}.
We will show this fact explicitly in Section~\ref{Gaussian}
for a bounded Gaussian distribution,
but it is true also for a Poisson distribution
with known background.
Therefore,
the Unified Approach
is the \emph{less powerful}
among new methods
in testing $\mu$ against the alternatives $\mu_1>\mu$,
which are more important near the boundary
than the alternatives $\mu_1<\mu$,
for which the Unified Approach has the highest power.
In other words,
the Unified Approach
is less sensitive than the other methods
to positive signals,
because small values of $\mu$
are more likely to be accepted
if the true value of $\mu$ is large.
It also produces upper limits that are too stringent
and unreliable from the physical point of view
\cite{Giunti-bo-99,Roe-Woodroofe-99,Mandelkern-Schultz-99,%
Giunti-clw2000,Astone-clw2000},
because it has too much power
in testing $\mu$ against the alternatives $\mu_1<\mu$
near the boundary.

\section{Gaussian distribution with boundary}
\label{Gaussian}

In order to illustrate the power of different methods,
let us consider an observable $x$
with Gaussian distribution around a
non-negative mean $\mu$ and
a standard deviation $\sigma$,
assumed to be known.
In this case $x$ is the estimator of $\mu$
($\hat\mu=x$)
and a measurement of $x$
gives a confidence interval for $\mu$,
which depend on the chosen method.

\begin{figure}
\begin{center}
\mbox{\includegraphics[bb=85 552 440 765,width=0.99\textwidth]{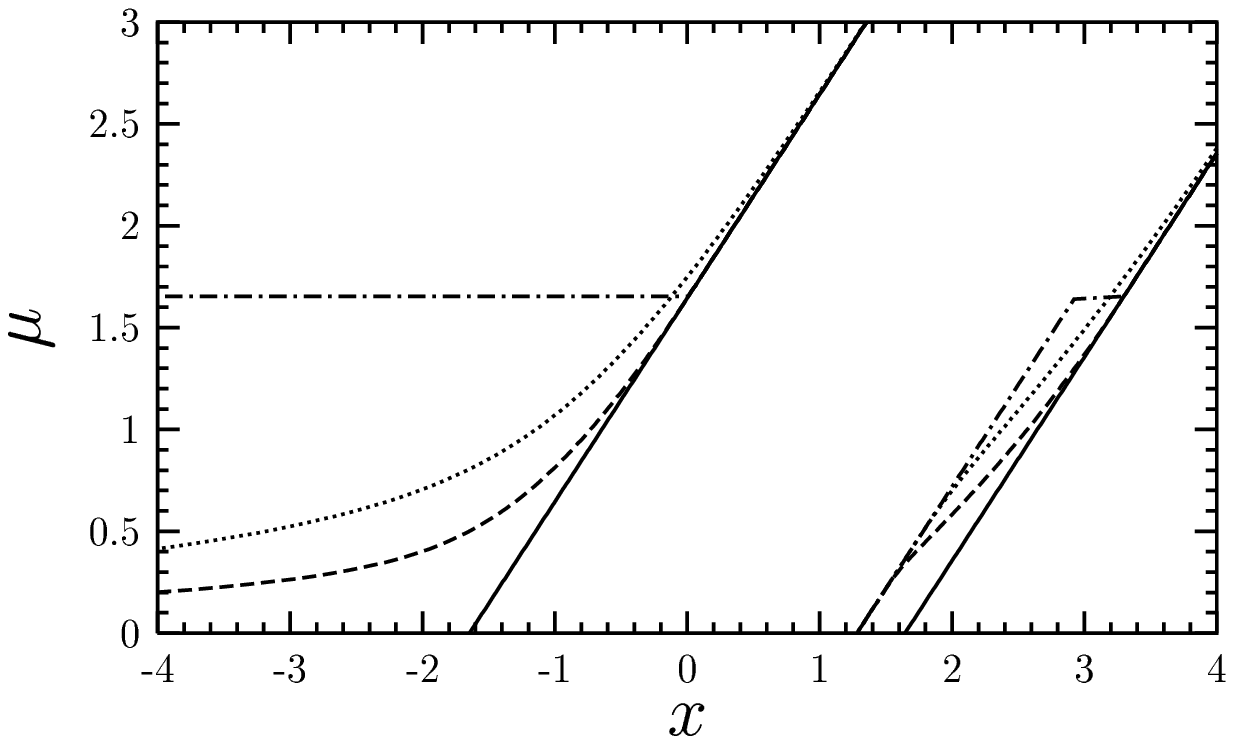}}
\end{center}
\caption{ \label{gauss-belt}
90\% CL confidence belts ($\alpha=0.10$)
in four methods
for an observable $x$
with Gaussian distribution around a
positive mean $\mu$ and
a standard deviation $\sigma=1$.
Solid lines:
Central Intervals;
Dashed lines:
Unified Approach \protect\cite{Feldman-Cousins-98};
Dotted lines:
Bayesian Ordering \protect\cite{Giunti-bo-99};
Dash-dotted lines:
Maximum Likelihood Estimator \protect\cite{Mandelkern-Schultz-99}.
}
\end{figure}

Figure~\ref{gauss-belt}
shows the 90\% CL confidence belts ($\alpha=0.10$)
for $\sigma=1$
corresponding to four different methods:
the standard
Central Intervals method and the
three new methods with correct coverage,
Unified Approach \cite{Feldman-Cousins-98},
Bayesian Ordering\footnote{
In Ref.~\cite{Giunti-bo-99}
the Bayesian Ordering method
is discussed explicitly only for
the case of a Poisson distribution with known background,
but,
as noted there,
the method can be generalized in a straightforward way
to other cases,
as that considered here.
} \cite{Giunti-bo-99}
and
Maximum Likelihood Estimator \cite{Mandelkern-Schultz-99}.
For $x\gg\sigma$
all the methods produce the same results,
far from the boundary $\mu\geq0$.
The Central Intervals
method gives an empty confidence interval
for $x\leq-1.64$.
The other three methods give non-empty
confidence intervals for any value of $x$,
which become upper limits for $x\leq1.28$.
For negative values of $x$
the Unified Approach give the most stringent
upper limits,
whereas the Maximum Likelihood Estimator
gives the upper limit
$\mu\leq1.65$
for any value of $x\leq0$.
At $\mu=1.65$,
the confidence belt
obtained with the
Maximum Likelihood Estimator method has discontinuous derivative
of the left edge at
$x=0$,
and a discontinuity of the right edge
from
$x=3.30$ to $x=2.92$.
The confidence belt
obtained with the Unified Approach
has discontinuous derivatives
of both left and right edges
at $\mu=1.65$,
where the left edge has $x=0$
and the confidence belt start to deviate
from the Central Intervals confidence belt
for decreasing $x$.
Both left and right edges
of the confidence belt
obtained with the Bayesian Ordering
are smooth for all values of $x$.

For small values of $\mu$
the acceptance intervals in the
Unified Approach,
Bayesian Ordering
and
Maximum Likelihood Estimator
are increasingly shifted to the left,
with respect to those in the Central Intervals method.
Hence,
among these methods,
the Maximum Likelihood Estimator
has highest power in testing small values of $\mu$ against larger alternatives,
followed by the Bayesian Ordering
and then by the Unified Approach.
The order of the power of the four methods
in testing small values of $\mu$ against smaller alternatives
is reversed.

\begin{figure}
\begin{center}
\mbox{\includegraphics[bb=55 335 485 760,width=0.99\textwidth]{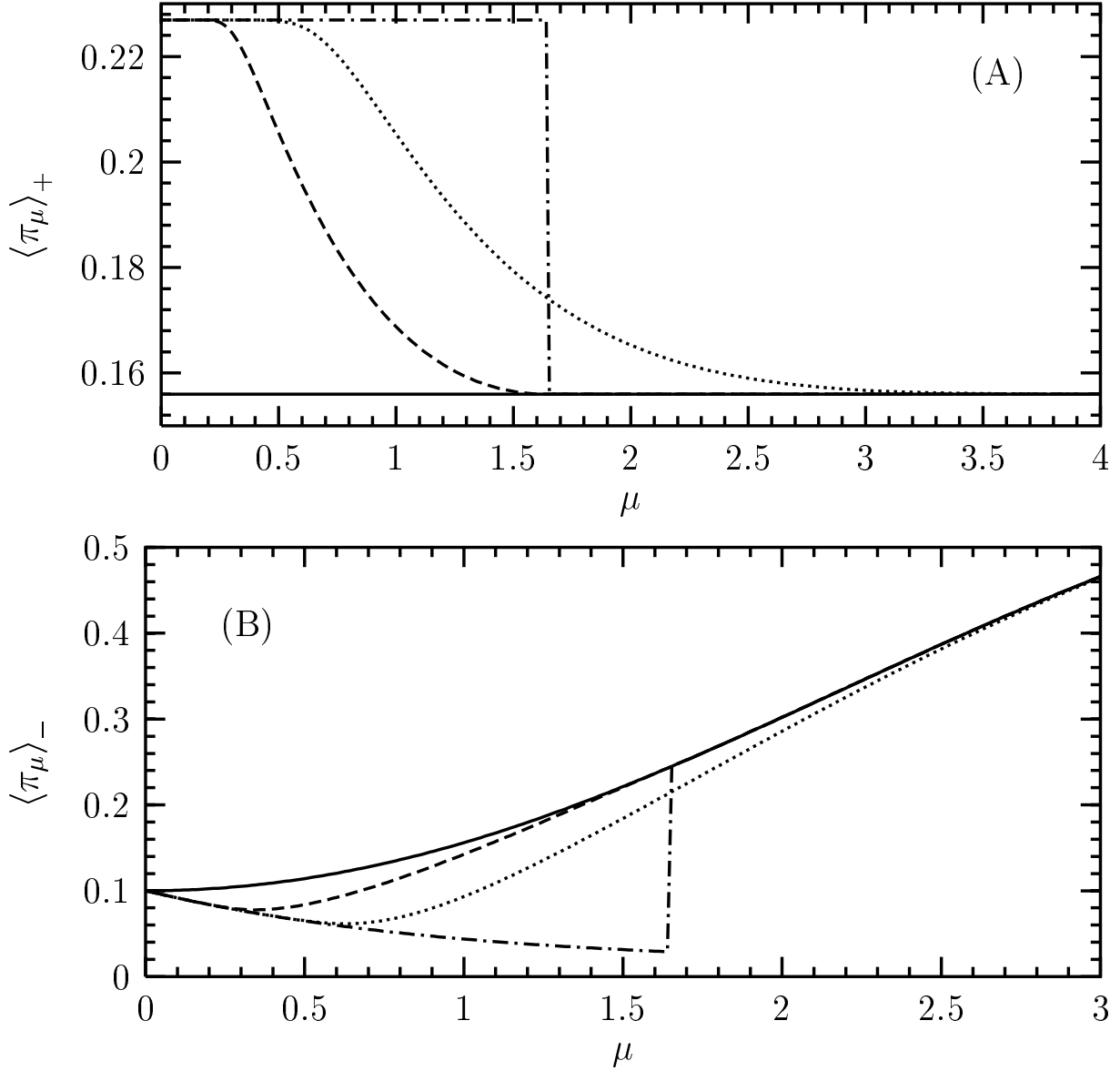}}
\end{center}
\caption{ \label{gauss-power}
Averaged power functions
of the
90\% CL confidence belts ($\alpha=0.10$)
in four methods
for a Gaussian distribution around a
positive mean $\mu$ and
a standard deviation $\sigma=1$.
(A): positive average power function
$\left\langle \pi_{\mu} \right\rangle_{+}$
with $\Delta\mu=1$
(see Eq.~(\ref{power-ave-plus}));
(B): negative average power function
$\left\langle \pi_{\mu} \right\rangle_{-}$
(see Eq.~(\ref{power-ave-minu})).
Solid line:
Central Intervals;
Dashed line:
Unified Approach \protect\cite{Feldman-Cousins-98};
Dotted line:
Bayesian Ordering \protect\cite{Giunti-bo-99};
Dash-dotted line:
Maximum Likelihood Estimator \protect\cite{Mandelkern-Schultz-99}.
}
\end{figure}

In order to give
a quantitative illustration of the power of the four methods
under consideration,
let us define the positive average power function
\begin{equation}
\left\langle \pi_{\mu} \right\rangle_{+}
=
\frac{1}{\Delta\mu}
\int_{\mu}^{\mu+\Delta\mu} \pi_{\mu}(\mu_1) \mathrm{d}\mu_1
\,,
\label{power-ave-plus}
\end{equation}
for an arbitrary $\Delta\mu$,
and
the negative average power function
\begin{equation}
\left\langle \pi_{\mu} \right\rangle_{-}
=
\frac{1}{\mu}
\int_{0}^{\mu} \pi_{\mu}(\mu_1) \mathrm{d}\mu_1
\,.
\label{power-ave-minu}
\end{equation}
These two functions,
for a Gaussian distribution with mean $\mu\geq0$,
standard deviation $\sigma=1$,
$\alpha=0.10$
and
$\Delta\mu=1$,
are plotted in Fig.~\ref{gauss-power}.

From Fig.~\ref{gauss-power}A
one can see that the Maximum Likelihood Estimator method
has the highest power
with respect to the alternatives $\mu_1>\mu$
if $\mu<1.65$.
The power of the Bayesian Ordering method
is higher than that of the Unified Approach
and both tend to the
constant power of the Central Intervals method
for large values of $\mu$.
All methods are unbiased
in testing larger alternatives
($\left\langle \pi_{\mu} \right\rangle_{+} > \alpha = 0.1$).

Figure~\ref{gauss-power}B
shows that the
order of the power of the four methods is reversed
when smaller alternatives are considered
and only the
Central Intervals method is unbiased.
The curves in Fig.~\ref{gauss-power}B
increase with increasing $\mu$
because the range $0 \leq \mu_1 < \mu$
increases with $\mu$
and values of $\mu_1$ far from $\mu$
lead to higher values of the power.
The value of $\left\langle \pi_{\mu} \right\rangle_{-}$
tends to $\alpha=0.10$ for $\mu\to0$
in all methods,
because by definition
$\pi_{\mu_0}(\mu_1)\to\alpha$ for $\mu_1\to\mu_0$
and the interval of integration in Eq.~(\ref{power-ave-minu})
shrinks to zero for $\mu\to0$.
As noted in Section~\ref{boundary},
a small power in testing smaller alternatives
is desirable in order to obtain
physically reliable upper limits.

Let us emphasize that the arbitrary definition
of the quantities
in Eqs.(\ref{power-ave-plus}) and (\ref{power-ave-minu})
is irrelevant for the quality of our conclusions.
Choosing other appropriate quantities
one always obtains the same classification
(Central Intervals,
Unified Approach,
Bayesian Ordering,
Maximum Likelihood Estimator)
of the four considered methods
in order of increasing power to test larger alternatives
and decreasing power to test smaller alternatives.

\section{Poisson distribution with background}
\label{Poisson}

In this section we discuss
as another example the case of
a Poisson distribution of counts $n$ with mean signal $\mu$ and
known background $b=5.2$.
We consider the same four methods
already considered in the previous section:
Central Intervals,
Unified Approach \cite{Feldman-Cousins-98},
Bayesian Ordering \cite{Giunti-bo-99}
and
Maximum Likelihood Estimator \cite{Mandelkern-Schultz-99}.
Since the discreteness of $n$ does not allow to
construct exact Central Intervals,
we follow the prescription described in Ref.~\cite{Mandelkern-Schultz-99}.

\begin{figure}
\begin{center}
\mbox{\includegraphics[bb=85 552 440 765,width=0.99\textwidth]{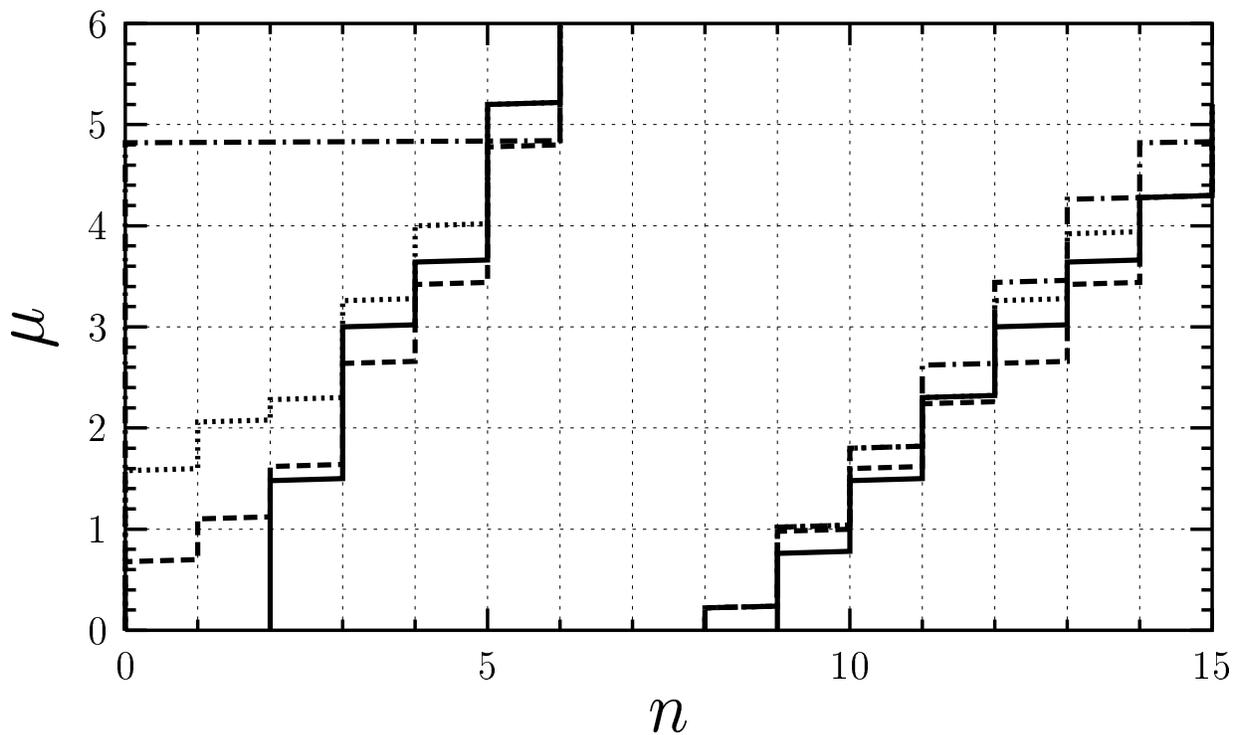}}
\end{center}
\caption{ \label{poisson-belt}
90\% CL confidence belts ($\alpha=0.10$)
in four methods
for a Poisson distribution of counts $n$ with mean signal $\mu$ and
known background $b=5.2$.
Solid lines:
Central Intervals;
Dashed lines:
Unified Approach \protect\cite{Feldman-Cousins-98};
Dotted lines:
Bayesian Ordering \protect\cite{Giunti-bo-99};
Dash-dotted lines:
Maximum Likelihood Estimator \protect\cite{Mandelkern-Schultz-99}.
}
\end{figure}

Figure~\ref{poisson-belt}
shows the 90\% CL confidence belts
in the four methods.
One can see that for $n\leq6$
the Maximum Likelihood Estimator
gives a constant upper limit $\mu\leq4.8$,
the Bayesian Ordering and Unified approach methods
give upper limits that decrease with $n$,
and the Central Intervals method gives an empty confidence interval
for $n<2$.

\begin{figure}
\begin{center}
\mbox{\includegraphics[bb=85 552 440 765,width=0.99\textwidth]{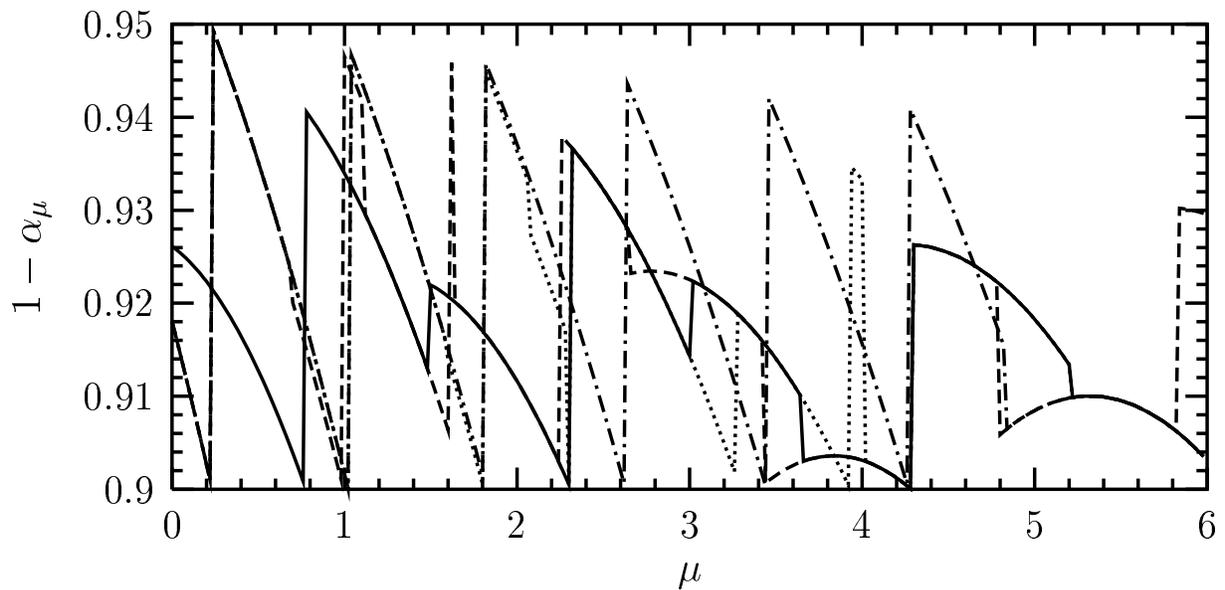}}
\end{center}
\caption{ \label{poisson-alp}
Integral probability $1-\alpha_\mu$
of the acceptance intervals of the
90\% CL confidence belts ($\alpha=0.10$)
in four methods
for a Poisson distribution of counts $n$ with mean signal $\mu$ and
known background $b=5.2$.
Solid line:
Central Intervals;
Dashed line:
Unified Approach \protect\cite{Feldman-Cousins-98};
Dotted line:
Bayesian Ordering \protect\cite{Giunti-bo-99};
Dash-dotted line:
Maximum Likelihood Estimator \protect\cite{Mandelkern-Schultz-99}.
}
\end{figure}

Because of the discreteness of $n$,
the acceptance intervals do not have exact integral probability
$1-\alpha$.
Their integral probability $1-\alpha_\mu$
is shown in
Fig.~\ref{poisson-alp}
as a function of $\mu$
in the four considered methods.
One can see that the dependence of $1-\alpha_\mu$ from
$\mu$ is very wild.
As a consequence,
also the average power functions
in Eqs.~(\ref{power-ave-plus}) and (\ref{power-ave-minu})
are wild functions of $\mu$.
In order to obtain smoother functions of $\mu$,
whose behaviour is not too difficult to be interpreted,
we consider the ratios
$\left\langle \pi_{\mu} \right\rangle_{+}/\alpha_\mu$
and
$\left\langle \pi_{\mu} \right\rangle_{-}/\alpha_\mu$.
These ratios are appropriate to evidence the presence of a bias,
which manifest itself when a ratio becomes less then one.

\begin{figure}
\begin{center}
\mbox{\includegraphics[bb=55 335 485 760,width=0.99\textwidth]{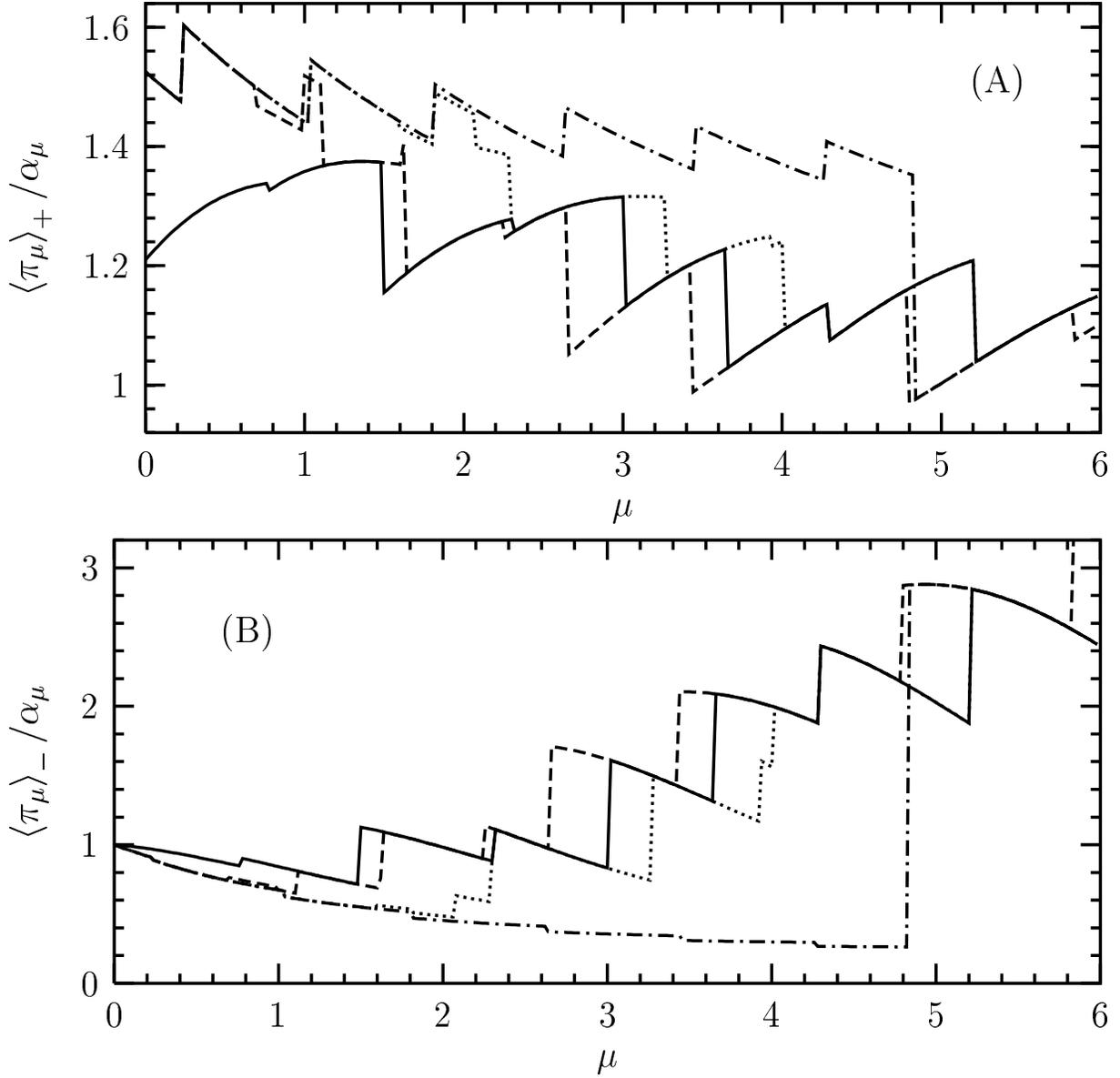}}
\end{center}
\caption{ \label{poisson-power}
Normalized average power functions
of the
90\% CL confidence belts ($\alpha=0.10$)
in four methods
for a Poisson distribution of counts $n$ with mean signal $\mu$ and
known background $b=5.2$.
(A): normalized positive average power function
$\left\langle \pi_{\mu} \right\rangle_{+}/\alpha_\mu$
with $\Delta\mu=1$
(see Eq.~(\ref{power-ave-plus}));
(B): normalized negative average power function
$\left\langle \pi_{\mu} \right\rangle_{-}/\alpha_\mu$
(see Eq.~(\ref{power-ave-minu})).
Solid line:
Central Intervals;
Dashed line:
Unified Approach \protect\cite{Feldman-Cousins-98};
Dotted line:
Bayesian Ordering \protect\cite{Giunti-bo-99};
Dash-dotted line:
Maximum Likelihood Estimator \protect\cite{Mandelkern-Schultz-99}.
}
\end{figure}

Figure~\ref{poisson-power}A
shows
the ratio
$\left\langle \pi_{\mu} \right\rangle_{+}/\alpha_\mu$,
with $\Delta\mu=1$
(see Eq.~(\ref{power-ave-plus})).
In spite of the precaution to divide
$\left\langle \pi_{\mu} \right\rangle_{+}$
by
$\alpha_\mu$,
the curves still have wild jumps because of the discreteness of $n$.
From Fig.~\ref{poisson-power}A
one can see that for most small values of $\mu$
the Maximum Likelihood Estimator
has the highest power in testing $\mu$
against larger alternatives,
followed in order by
Bayesian Ordering,
Unified Approach
and Central Intervals.
The Unified Approach has higher (or equal) power than Central Intervals
for $\mu\lesssim2.6$,
and smaller (or equal) power mfor highest values of $\mu$.

Figure~\ref{poisson-power}B
shows
the ratio
$\left\langle \pi_{\mu} \right\rangle_{-}/\alpha_\mu$
(see Eq.~(\ref{power-ave-minu})).
One can see that even the Central Intervals method is slightly biased.
This is due to the fact that,
because of the discreteness of $n$,
it is not possible to construct exactly central acceptance intervals.
Fig.~\ref{poisson-power}B
shows that
Central Intervals and Unified Approach have the highest power
in testing $\mu$ against smaller alternatives
(Central Intervals for $\mu\lesssim2.2$
and Unified Approach
for highest values of $\mu$).
The Maximum Likelihood Estimator method
has the smallest power
for $\mu\leq4.8$.

In conclusion of this section,
we have shown with a specific example that also in the case
of a Poisson process with background,
among the recently proposed methods with correct coverage,
Maximum Likelihood Estimator yields confidence intervals with
the highest physical significance
(with the criteria discussed in Section~\ref{boundary}),
followed in order by
Bayesian Ordering and Unified Approach.

\section{Conclusions}
\label{Conclusions}

In conclusion,
we have considered
the power of Frequentist methods,
which quantifies their capability
to avoid Type II errors.
We have connected the power with the physical significance
of confidence intervals,
\textit{i.e.}
their degree of reliability.

Considering the case of a parameter bounded to be non-negative,
we have shown that near the boundary
a (biased) method that
has
large power in testing the parameter against larger alternatives
and
small power in testing the parameter against smaller alternatives
produces confidence intervals (upper limits)
that are physically more significant.

We have shown that
among the recently proposed Frequentist methods
with correct coverage
(Unified Approach \cite{Feldman-Cousins-98},
Bayesian Ordering \cite{Giunti-bo-99},
Maximum Likelihood Estimator \cite{Mandelkern-Schultz-99}),
the widely used Unified Approach
yields upper limits with the smallest physical significance.
The upper limits with the highest physical significance
are produced by the Maximum Likelihood Estimator method.

We have illustrated our arguments in
the cases of a bounded Gaussian distribution
(Section~\ref{Gaussian})
and a Poisson distribution with known background
(Section~\ref{Poisson}).

\newpage

\end{document}